\documentclass[1p]{elsarticle}

\usepackage{lineno,hyperref}

\modulolinenumbers[5]

\journal{Paleoclimate}

\biboptions{authoryear}

\begin{document}

\begin{frontmatter}

\title{Millennial-scale stable oscillations between sea ice and convective deep water formation}


\author{Raj Saha \corref{mycorrespondingauthor}}
\address{University of Minnesota, Department of Mathematics \\Mathematics and Climate Research Network}
\cortext[mycorrespondingauthor]{Corresponding author}
\ead{rajsaha80@gmail.com}

\begin{abstract}
During the last ice age there were several quasi-periodic abrupt warming events. The climatic effects of the so-called Dansgaard-Oeschger (DO) events were felt globally, although the North Atlantic experienced the largest and most abrupt temperature anomalies. Similar but weaker oscillations also took place during the interglacial period. This paper proposes an auto-oscillatory mechanism between sea ice and convective deep water formation in the north Atlantic as the source of the persistent cycles. A simple dynamical model is constructed by coupling and slightly modifying two existing models of ocean circulation and sea ice. The model exhibits mixed mode oscillations, consisting of decadal scale small amplitude oscillations, and a large amplitude relaxation fluctuation. The decadal oscillations occur due to the insulating effect of sea ice and leads to periodic ventilation of heat from the polar ocean. Gradually an instability builds up in the polar column and results in an abrupt initiation of convection and polar warming. The unstable convective state relaxes back to the small amplitude oscillations from where the process repeats in a self-sustained manner. Freshwater pulses mimicking Heinrich events cause the oscillations to be grouped into packets of progressively weaker fluctuations, as also observed in the proxy records. Modulation of this stable oscillation mechanism by freshwater and insolation variations could account for the distribution and pacing of DO and Bond events. Physical aspects of the system such as sea ice extent and oceanic advective flow rates could determine the characteristic 1,500 year timescale of DO events. 
\end{abstract}

\begin{keyword}
Dansgaard-Oeschger events; Heinrich events; sea ice; deep water formation; climate oscillations; box model
\end{keyword}

\end{frontmatter}


\section{Introduction}

\begin{figure}[htbp]
\begin{center}
	\includegraphics[width=5in]{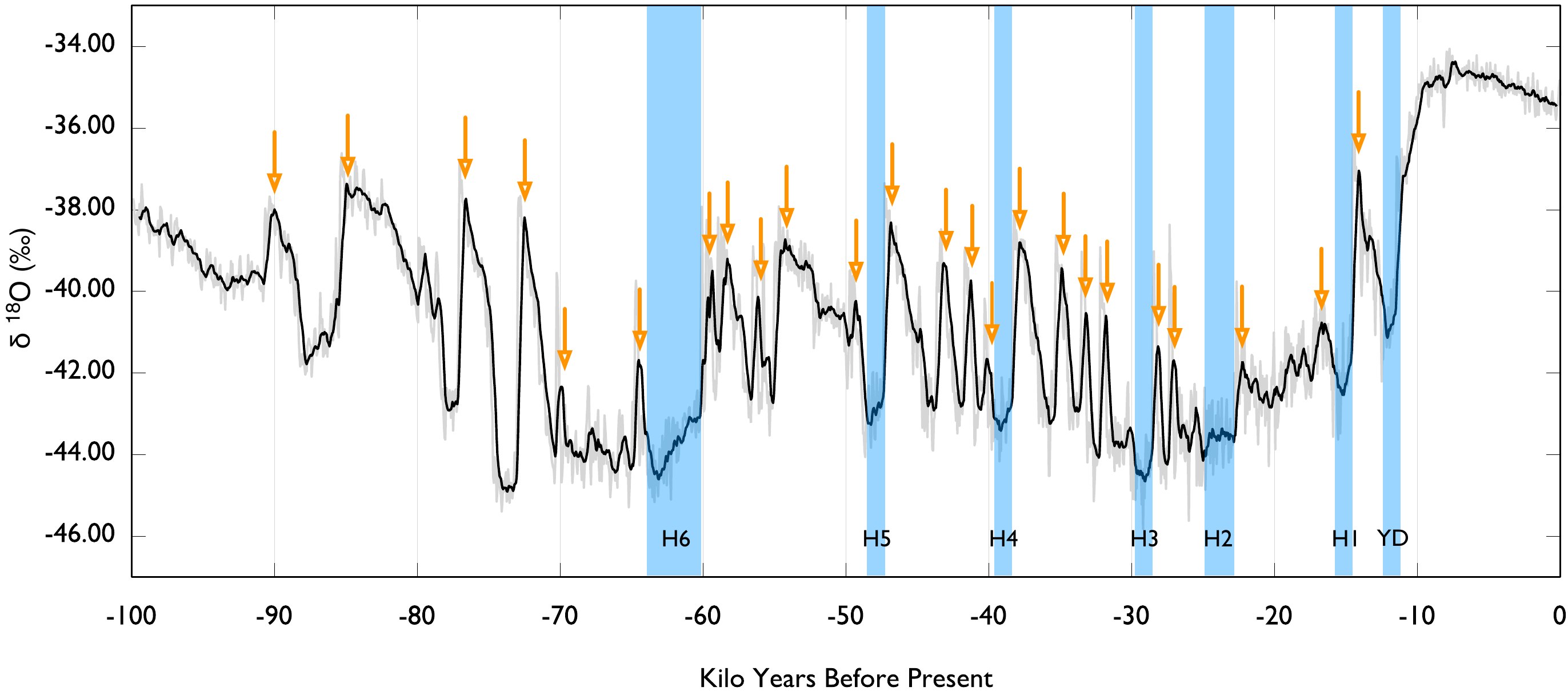}
\caption{
Oxygen-18 isotopic concentrations in Greenland ice cores (NGRIP) showing the variation of near surface air temperatures over the last 100,000 years. Orange arrows mark individual DO events and the blue vertical stripes mark Heinrich events numbered 1 through 6, and the Younger Dryas flooding event.
}
\label{fig:DO}
\end{center}
\end{figure}

Climate proxy data from ice \citep{dansgaard1993} and sediment \citep{wang2001} cores reveal that the climate of the north Atlantic underwent large quasi periodic fluctuations during the last glacial period (figure \ref{fig:DO}). These are the so-called Dansgaard-Oeschger (DO) events which occurred in intervals of about 1,500 years. Fluctuations on a similar timescale but with lesser intensity are also seen during the Holocene warm period, known as Bond events \citep{bond1997}. This paper proposes an auto-oscillatory mechanism between sea ice and convection in the north Atlantic to be the driving source of the pervasive millennial-scale cycles. A coupled sea ice-ocean box model is used to show the existence of stable oscillations whose pacing can be modulated by glacial meltwater pulses to produce temporal patterns very similar to the paleo-proxy record. 

The DO events were characterized by an abrupt warming on a decadal scale followed by slower cooling. These abrupt warming events were felt most strongly in the north Atlantic. Proxy records from Greenland \citep{ngrip} and Antarctica \citep{petit1999} show that the temperature anomalies in the northern and southern hemispheres were out of phase by about 800 to 1,200 years, the magnitude of anomalies being much smaller over Antarctica. This `bipolar seesaw' behavior suggests a global reorganization of heat flow on the surface via changes in oceanic circulation \citep{stocker2003}. If the total amount of heat in the climate system is conserved then warming in one region must be balanced by cooling elsewhere on timescales comparable to the transit time of the global oceanic circulation \citep{toggweiler2001}. 

The warming over Greenland was also more abrupt than elsewhere, suggesting that the source of the heat anomaly was located in the north Atlantic. At present, a region of convective deep water formation exists in the Norwegian seas. The cooling and subsequent sinking of surface waters through buoyant instability provides a mechanism for the deep ocean to ventilate heat to the atmosphere. Freshwater perturbations via sudden bursts of glacial meltwater, i.e. Heinrich events, could disrupt the convective process and abruptly alter the overturning circulation and climate state \citep{rahmstorf_2002,hofmann_rahmstorf_2009}. 

Heinrich events preceded some, but not all DO events. While there is little doubt that a large freshwater perturbation can disrupt and destabilize the convective process, it is not clear how a single pulse of freshwater could trigger multiple DO cycles. There is no evidence either to suggest that each DO event was triggered by a separate meltwater pulse. The timescale for ice sheet instability in models \citep{alley_1994} is also significantly greater than the pacing of DO events. Thus Heinrich events cannot be the driver of DO events. 

A number of hypotheses have been proposed to explain the source of oscillations. Some propose astronomical forcing through variations in solar output \citep{braun2005} or tidal resonance from the moon's orbit \citep{keeling2000}. Other studies suggest internal climatic mechanisms whereby steady salinity forcing, salting at low latitudes and freshening at high, can produce self-sustained oscillations \citep{winton_1993,sakai_peltier_1995,sakai_peltier_1999, lenderink1994,haarsma_2001, verdiere2006}. These mechanisms do not explain many of the observed characteristics of DO events. For example, proxies of solar output do not show correlation with the climatological anomalies from DO events \citep{muscheler_2006}. The lunar hypothesis, which connects tidal resonances to the destabilization of ice sheets and consequent disruption of the overturning circulation, do not explain the variability in the pacing of DO events through time. Internal thermohaline oscillations are unstable to freshwater perturbations and are unlikely to have persisted through the glacial and Holocene epochs and through large variations in surface freshwater conditions. A stable oscillation mechanism is needed that is robust against freshwater anomalies and glacial-interglacial scale changes in the background climate.

In this paper a sea ice-convection oscillator is proposed as the pacemaker for DO events. Some previous studies have noted the importance of sea ice in initiating polar convection \citep{haarsma_2001,timmermann_2003}. A Van der Pol type oscillator was proposed by Saltzman  \citep{saltzman2002} in which sea ice cover regulates the positive feedback from oceanic uptake of CO$_2$ to produce relaxation oscillations. Although atmospheric CO$_2$ variations were unlikely to be the cause for DO events, as the convective region is not large enough to account for the required CO$_2$ flux, sea ice could in a similar manner regulate the heat content of the deep ocean \citep{rial_yang_2007}. This insulating property of sea ice, its on/off presence, and geophysical constraints on its spatial extent could make it a suitable component of a stable oscillation mechanism. The following sections describe a simple dynamical model composed of sea ice and an ocean box model which interact to produce stable millennial scale oscillations. The simplified modeling approach makes it possible to capture the minimum number of processes that are necessary to produce oscillations and to analyze the model's dynamical behavior.

\section{A simple dynamical model}
\label{sec:model}

The model consists of an ocean box model \citep{verdiere2006} coupled to a thermodynamic sea ice model \citep{GT2001b}. The two components are among of the simplest available models of the respective processes.  In the ocean model there are three meridional boxes arranged in two vertical layers. Each box has a uniform temperature and salinity which can change dynamically. Advective transport between adjacent boxes is driven by the horizontal pressure in the top layer and the net flow in the model is conserved. Sea ice forms or melts on the top polar box if the temperature goes below or above the melting point. The dominating influence of sea ice on the ocean circulation is through the regulation of heat exchange between the top polar box and the atmosphere. There is no dynamic atmosphere, just a static temperature gradient applied to the top layer boxes.

The following modifications are made to the original models:

\begin{itemize}
	\item The masses of individual ocean boxes are computed run-time and not kept fixed as in \citep{verdiere2006}. This is done to avoid an unphysical density anomaly arising out of the convective mechanism. The mixing of two equal density water masses produces a slightly higher density of the combined water mass and this one time-step anomaly destabilizes the system to produce the reported `free oscillations'. This scenario is an artifact of the discretized description of the model and not due to a physical process. With the mass correction, the sea ice decoupled ocean model does not exhibit oscillations. While there is more than one way to resolve this computational issue, such as by increasing the convective mixing time-step or by using higher order non-linear terms in the equation of state, updating the mass of the boxes run-time is the least invasive of these ways in preserving the fundamental aspects of the original models. 
	
	\item Diffusive mixing is not considered as its inclusion only affects the response timescale and not the qualitative character of the model's behavior. 

	\item The extent of the southernmost box meridional zone is shortened to 45$^{\circ}$- 60$^{\circ}$ N to only consider the northern arm of the overturning circulation. 
	
	\item The distribution fractions for precipitation is different owing to the altered geometry of the system.  
			
	\item Sea ice is assumed to be a perfect insulator, unlike \citep{verdiere2006} where it had non zero heat permeability.
	
	\item Heat and salinity adjustments due to brine rejection and latent heat of fusion of ice is ignored. The contribution of these terms in the governing equations is effectively zero in one cycle of sea ice advance and retreat and their exclusion does not alter the qualitative character of the model. These terms are left out so that only the  ingredients necessary for oscillations can be examined.
\end{itemize}

A description of the basic set up of the model is given below which includes the modifications to the original components. The default parameter values and units are listed in Table \ref{table:parameters}. 

\begin{table}[h!]
\begin{center}
\begin{tabular}{ r | l | l  }
\hline \hline
$g$ & Acceleration due to gravity & 9.8 m s$^{-2}$\\
$C_p$ & Specific heat capacity of water & 4000 J kg$^{-1}$ K$^{-1}$\\
$\rho_{o}$  & Reference density of sea water & 1028 kg m$^{-3}$\\
$\rho_{\mathrm{ice}}$  & Density of ice & 920 kg m$^{-3}$\\
$\alpha$ & Thermal expansion coefficient & 1.7$\times 10^{-4}$ K$^{-1}$\\
$\beta$ & Haline contraction coefficient & 0.76 \\
$T_0$ & Freezing point of water & 273 K\\
$L_{f}$ & Latent heat fusion of ice & 3.34$\times10^{5}$ J kg$^{-1}$\\
$\lambda$ & Heat exchange coefficient between atmosphere and ocean & 30 W m$^{-2}$ K$^{-1}$\\
$T_{r}$ & Reference temperature & 283 K\\
$S_{r}$ & Reference salinity & 0.035\\
$\Delta t$ & Integration time step & 0.25 years\\
& \\
$\eta$ & Thermal gradient & -0.74 K/$^{\circ}$ lat \\
$\epsilon$ & Net evaporation from surface tropical box & 12 mm/day \\
$q_{i}$ & Salinity forcing ratios for surface boxes & +0.3, -0.2, -0.1 \\
$T^a_1$ & Atmospheric temperature above box 1 & 297 K\\
$C$ & Advective transport coefficient (inverse of friction coefficient) & $10^{-5}$ m$^{2}$ s kg$^{-1}$\\
$d_{\mathrm{ice}}$ & Thickness of sea ice & 2 m\\
$\tau$ & growth-melt timescale of sea ice& 3 years \\
\hline
\end{tabular}
\label{table:parameters}
\caption{Description of model parameters.}
\end{center}	
\end{table}	

\subsection{Geometry}
\begin{figure}[htbp]
\begin{center}
	\includegraphics[width=3.5in]{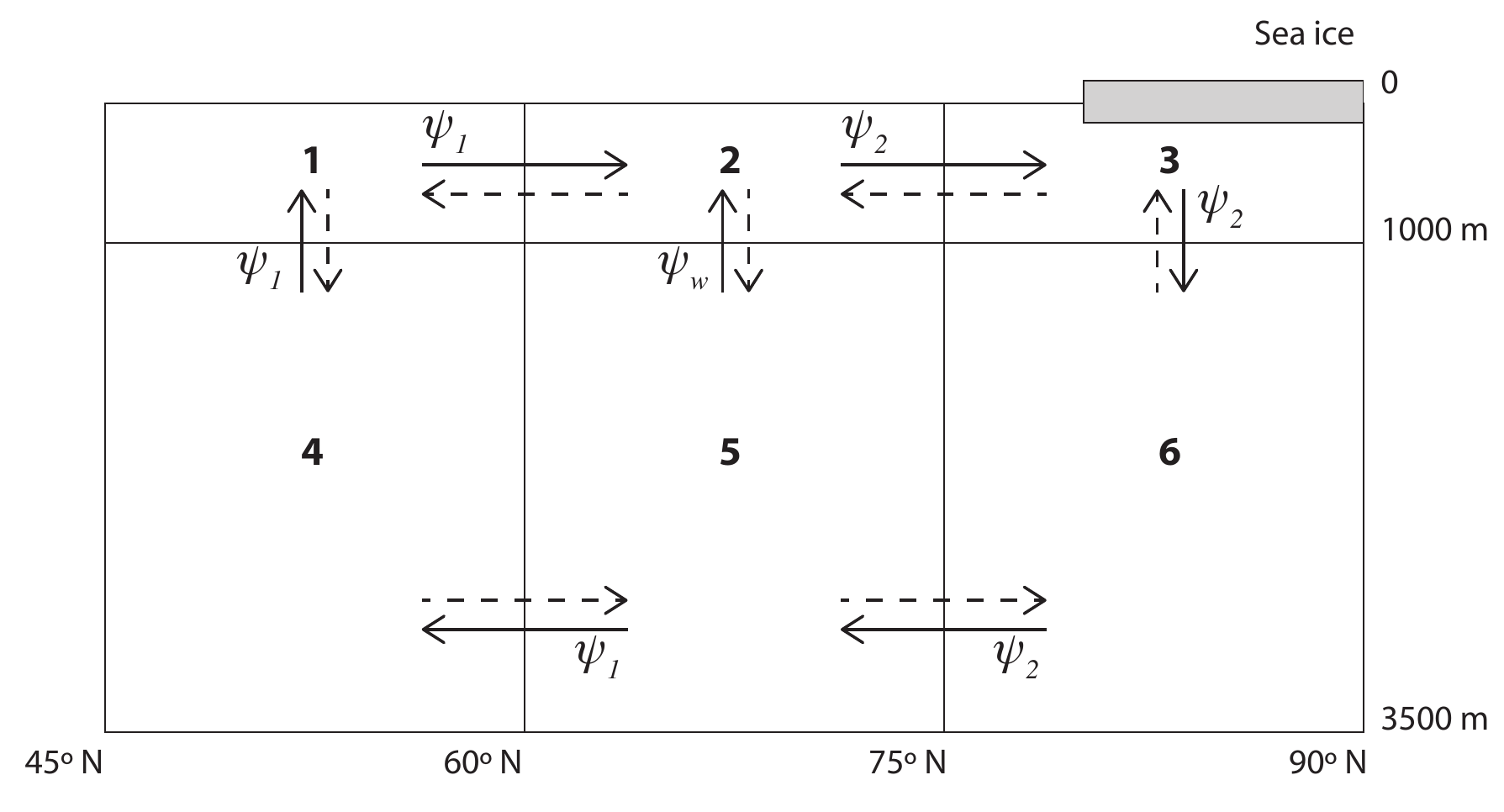}
\caption{Schematic of the box model. The solid and dashed arrows indicate up and downstream flows respectively. The height and width of the boxes are shown to scale, however the volumes of boxes is not to scale with the areas as shown due to the three dimensional geometry of the system.}
\label{fig:box-schematic}
\end{center}
\end{figure}

The ocean boxes are arranged in three meridional zones and two vertical layers. The same indexing scheme is used as in \citep{verdiere2006}. Areas of the boxes are computed by treating the entire ocean as part of a shell enveloping a sphere of the Earth's radius. Flows take place between boxes sharing a common face. The extent of sea ice is limited to the top polar box only. A schematic of the model geometry with the box indexing scheme is shown in figure \ref{fig:box-schematic}. 

\subsection{Governing equations}
The system is defined by a set of ordinary differential equations accounting for the rate of change of heat and salt content of each box, and the rate of change of sea ice extent. 
\begin{eqnarray} \label{eq:goveq}
	m_i C_{\mathrm{p}} \frac{dT_i}{dt}&=& F^T_i + \rho_o C_{\mathrm{p}} \psi_{i,j} T_j + \xi^T_i  \label{eq:goveq-T}\\
	m_i \frac{dS_i}{dt}&=& F^S_i + \rho_0\psi_{i,j} S_j + \xi^S_i \label{eq:goveq-S}\\
	\frac{df}{dt}&=&\frac{k_1(T_0-T_3)}{\tau}+k_2 f \label{eq:goveq-f}
\end{eqnarray}
Here $i$ is the index of a box. Terms on the right hand-side of equations (\ref{eq:goveq-T}) and (\ref{eq:goveq-S}) are contributions to heat and salinity fluxes via atmospheric forcing ($F^T_i, F^S_i$), advective flows ($\psi_{i,j}$), and convective mixing ($\xi^T_i, \xi^S_i$). The flow terms, $\psi_{i,j}$, represent flows between boxes $i$ and $j$, and $\psi_{i,j}T_j$ and $\psi_{i,j}S_j$ are implied summations over $j$ to account for the net flux of heat and salinity into box $i$ from its adjoining boxes. The convective mixing terms are temperature and salinity adjustments on uniformly mixing the top and bottom boxes in a column. 

In equation (\ref{eq:goveq-f}) $f$ represents the fractional sea ice extent, $k_1$ is the fractional  growth-melt rate constant of sea ice per unit difference in temperature from freezing, $\tau$ is the timescale for sea ice growth-melt, and $k_2$ accounts for precipitation on existing sea ice. Sea ice also cuts off heat and salinity exchange between the top polar box and the atmosphere in proportion to $f$. 

\subsection{Thermal forcing}
Each of the top layer boxes are made to relax to a static atmospheric temperature, with the air temperatures decreasing linearly with latitude. The heat exchange rate for boxes 1 and 2 are
\begin{equation}
	F^T_i = \lambda (T^a_i-T_i)A_i
\end{equation}
where $A_i$ is the surface area of the box in contact with the atmosphere, $T^a_i$ is the atmospheric temperature, and $\lambda$ is the heat exchange coefficient. For the polar box the area of contact is adjusted according to the fraction of area covered by sea ice
\begin{equation}
	F^T_3 = \lambda (T^a_3-T_3)A_3(1-f)
\end{equation}

The atmospheric temperatures are prescribed by the thermal gradient parameter $\eta$, and a prescribed $T^a_1$. 
\begin{equation}
	T^a_i = \eta\, (\theta_i -\theta_1)+ T^a_1
\end{equation}
where $\theta_i$ is the middle latitude of box $i$. 

\subsection{Salinity (freshwater) forcing}
In addition to thermal forcing from the atmosphere, each top layer box receives salinity forcing due to evaporation or precipitation. Salinity flux from the atmosphere is positive for net evaporation, as in box 1, and negative for net precipitation in the other two boxes. The addition of salinity is such that the total amount of salt is conserved in the system. The distribution fractions of precipitation is different from \citep{verdiere2006} to scale for evaporation and precipitation in the absence of a tropical ocean box. The distribution fractions are
\begin{eqnarray}
	q_1&=&+0.3\\
	q_2&=&-0.2\\
	q_3&=&-0.1
\end{eqnarray}
The salinity flux terms for each top layer box is given by
\begin{eqnarray}
	F^S_i&=&\rho_o S_o \epsilon A_1 q_i, i=\{1,2\} \\
	F^S_3&=&\rho_o S_o \epsilon A_1 (1-f)q_3
\end{eqnarray}
where $\rho_0$ and $S_0$ are reference density and salinity, $\epsilon$ is the net evaporation from box 1 in mm/day.

\subsection{Advection}
Advective flows in the model are driven by horizontal pressure gradients given by
\begin{equation}
	\psi_i = \psi_{i,i+1}=C A_{i,i+1}(P_i-P_{i+1}), i=\{1,2\}
\end{equation}
where the parameter C is the inverse of a friction coefficient whose value chosen to give appropriate flow rates for the North Atlantic, $A_{i,i+1}$ is the area of contact between adjoining boxes, and $P_i, P_{i+1}$ are the respective pressures in the boxes. Pressure is formulated in such a way that the net pressure integrated along a vertical column is zero. 
\begin{eqnarray}
	P_i&=&-\frac{1}{2} \frac{h_{i+3}}{h_{i}+h_{i+3}} (h_{i} \rho_{i}+h_{i+3} \rho_{i+3})g \\
	P_{i+3}&=&-\frac{h_{i}}{h_{i+3}} P_i
\end{eqnarray}
where $h_{i}$ and $h_{i+3}$ are the depths, and $\rho_{i}$ and $\rho_{i+3}$ are the densities of the top and bottom boxes in a vertical column. Density is computed from the linear equation of state given as
\begin{equation}
	\rho_{i}=\rho_o(1-\alpha (T_i-T_r)+\beta (S_i-S_r))
\end{equation}
where $\alpha$ and $\beta$ are thermal expansion and haline contraction coefficients respectively, and $T_r, S_r$ are reference temperature and salinity values. 

Volume conservation requires flows between any pair of top layer boxes to be matched by flow in the opposite direction in the bottom layer. The up/downwelling flux between boxes 2 and 5 is similarly derived from volume constraints as
\begin{equation}
	\psi_w =\psi_{5,2}=\psi_2-\psi_1
\end{equation}

A positive $\psi_i$ implies poleward flow on the surface, and a positive $\psi_w$ implies upwelling. 

\subsection{Convection}
The convective scheme mixes the heat and salt contents uniformly between the boxes in a vertical column when a buoyant instability, $\rho_i\geq\rho_{i+3}$, occurs. The equalized temperature and salinity values are given by
\begin{eqnarray}
	\bar{S}_i&=& \frac{m_i S_i+m_{i+3} S_{i+3}}{m_i+m_{i+3}}\\
	\bar{T}_i&=& \frac{m_i T_i+m_{i+3} T_{i+3}}{m_i+m_{i+3}}
\end{eqnarray}		
So that the heat and salinity adjustment terms in equations (\ref{eq:goveq-T}) and (\ref{eq:goveq-S}) are then
\begin{eqnarray}
	\xi^T_i&=&\frac{\bar{T_i}-T_i}{\Delta t} \\
	\xi^S_i&=&\frac{\bar{S_i}-S_i}{\Delta t} 
\end{eqnarray}		
where $\Delta t$ is the time step over which convection is carried out, which by default is equal to the integration time step. 

\subsection{Sea ice}
Sea ice forms/melts when the temperature of the surface polar box goes below/above freezing. The governing growth rate by volume \citep{GT2001b} is given by
\begin{equation}
	\frac{dV_\mathrm{ice}}{dt}=\frac{\rho_o C_{\mathrm{p}}V_3}{\rho_{\mathrm{ice}}L_f \tau}(T_0 - T_3) + f p_3
\end{equation}
where $V_3$ is the volume of the polar box, $\tau$ is a prescribed timescale for sea ice growth, $T_0$ is the freezing temperature, and $p_3$ is the precipitation rate by volume per unit time given by
\begin{equation}
	p_3=A_3 \epsilon |q_3|
\end{equation}
so that any precipitation on existing sea ice is turned into additional sea ice. The volume of ice formed is spread over the polar box with thickness $d_{\mathrm{ice}}$. The fractional growth rate in sea ice is then
\begin{eqnarray}
	\frac{df}{dt}&=&\frac{1}{d_{\mathrm{ice}} A_3}\frac{dV_\mathrm{ice}}{dt} \nonumber\\
	&=&\frac{1}{d_{\mathrm{ice}} A_3}\Bigg(\frac{\rho_o C_{\mathrm{p}}V_3}{\rho_{\mathrm{ice}}L_f \tau}(T_0 - T_3) + f p_3\Bigg)
\end{eqnarray}
which is expressed in a more compact form as equation (\ref{eq:goveq-f}). The constants $k_1$ and $k_2$ are then given by
\begin{eqnarray}
	k_1&=&\frac{1}{d_{\mathrm{ice}} A_3} \frac{\rho_o C_{\mathrm{p}}V_3}{\rho_{\mathrm{ice}}L_f } \\
	k_2&=&\frac{p_3}{d_{\mathrm{ice}}A_3}
\end{eqnarray}

As discussed earlier the only influence of sea ice in this model is the alteration of heat and salinity exchange of the polar box with the atmosphere. Brine rejection and latent heat of fusion are ignored as their contributions in one cycle of sea ice advance and retreat equal zero. Although those terms affect the timescale of oscillations, their exclusion does not change the qualitative character of the model and they are therefore ignored in order to focus on the driving ingredients only. 

\section{Model dynamics}

\begin{figure}[htbp]
\begin{center}
	\includegraphics[width=5in]{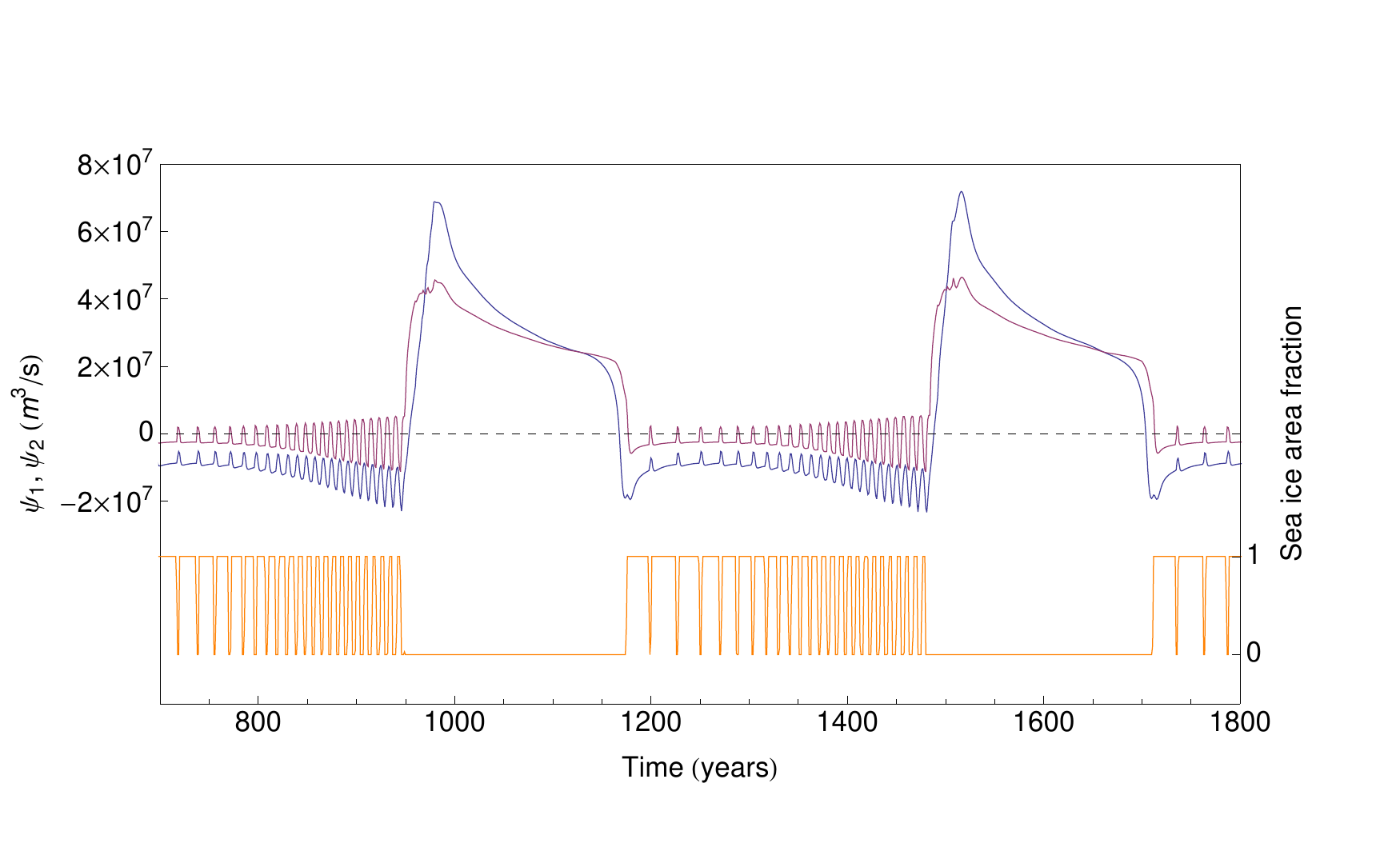}
\caption{Variation in advective fluxes $\psi_1$ (blue), $\psi_2$ (purple), and sea ice (orange) with time. Sea ice extent is typically either on or off. }
\label{fig:oscillations}
\end{center}
\end{figure}

The model exhibits two distinct equilibrium states and two self-sustained oscillating states. The steady states have opposite flow directions, termed as thermal (TH) for poleward flows in the top layer and haline (HA) in the opposite direction. One of the oscillating states (OS-2) is composed of two alternating modes, a small amplitude multi-decadal scale oscillation and a large millennial scale fluctuation (figure \ref{fig:oscillations}). The other  oscillating state (OS-1) is the small amplitude oscillation mode sustained indefinitely. 

\begin{figure}[htbp]
\begin{center}
	\includegraphics[width=5in]{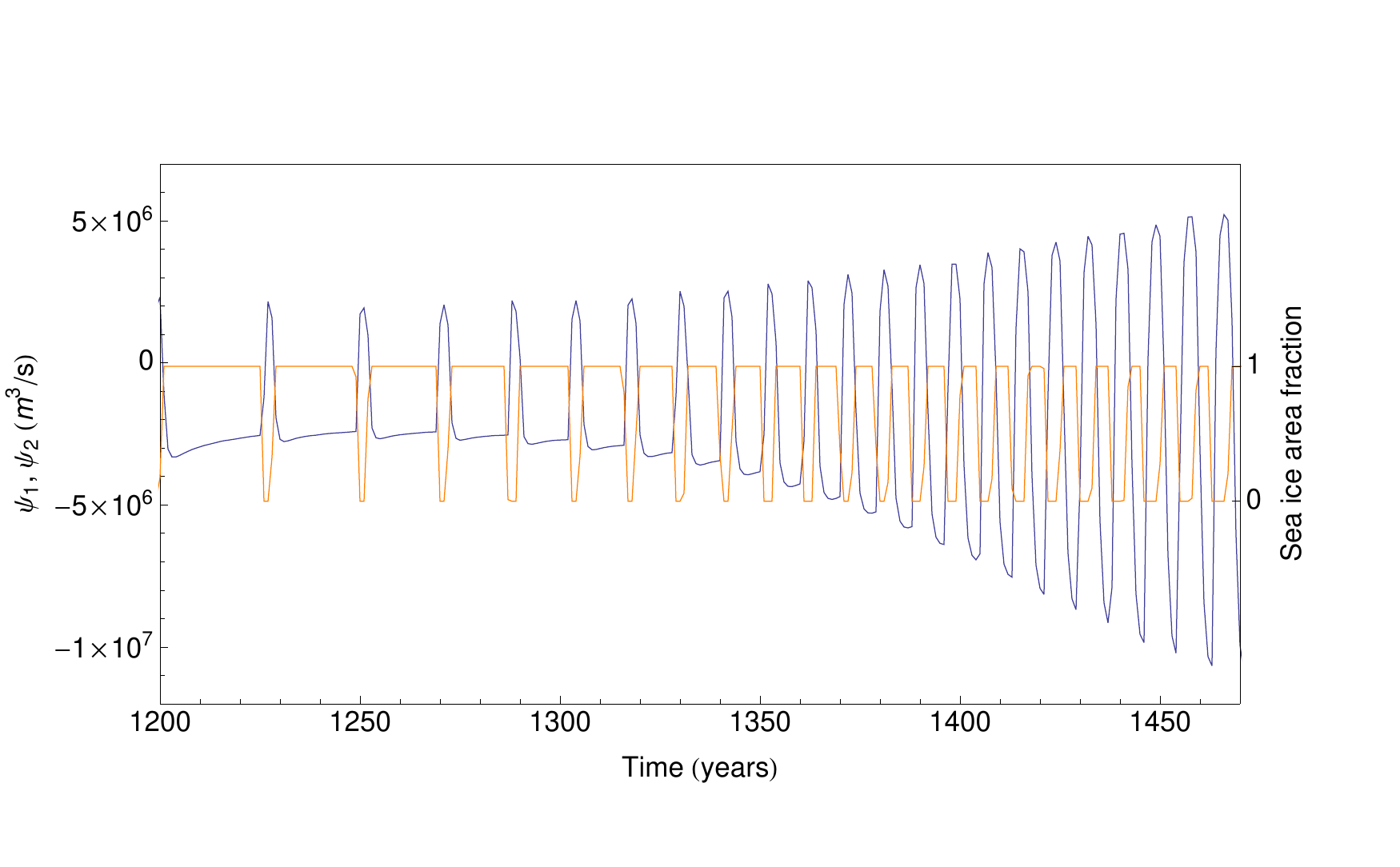}
\caption{Antagonistic variations in sea ice extent and $\psi_2$. These decadal scale oscillations are precursors to a convective instability in the polar water column. }
\label{fig:psi2_SI}
\end{center}
\end{figure}

The large amplitude component in oscillations is initiated when the density of the top layer polar box exceeds the density of the bottom box. Convection occurs when the instability condition is met and it results in an abrupt increase in poleward advective flux and with it an abrupt polar warming. The convective state is transient following which the system relaxes back to the small amplitude oscillations. During this phase sea ice and the polar advective flux, $\psi_2$, oscillate antagonistically with the advective flux switching between positive and negative, figure (\ref{fig:psi2_SI}). The amplitude of these oscillations increase with time until a convective instability is reached in the polar column. 

In each phase of sea ice advance and retreat the top polar box loses net heat. This is in spite of the insulating effect of sea ice. While sea ice covers the polar box its temperature increases until this causes the ice to melt and retreat. The lowering of density due to the temperature increase drives the advective flux poleward, and poleward heat transport. The high temperatures also mean higher ocean-atmosphere temperature gradient, which results in a greater loss of heat. 

\begin{figure}[htbp]
\begin{center}
	\includegraphics[width=5in]{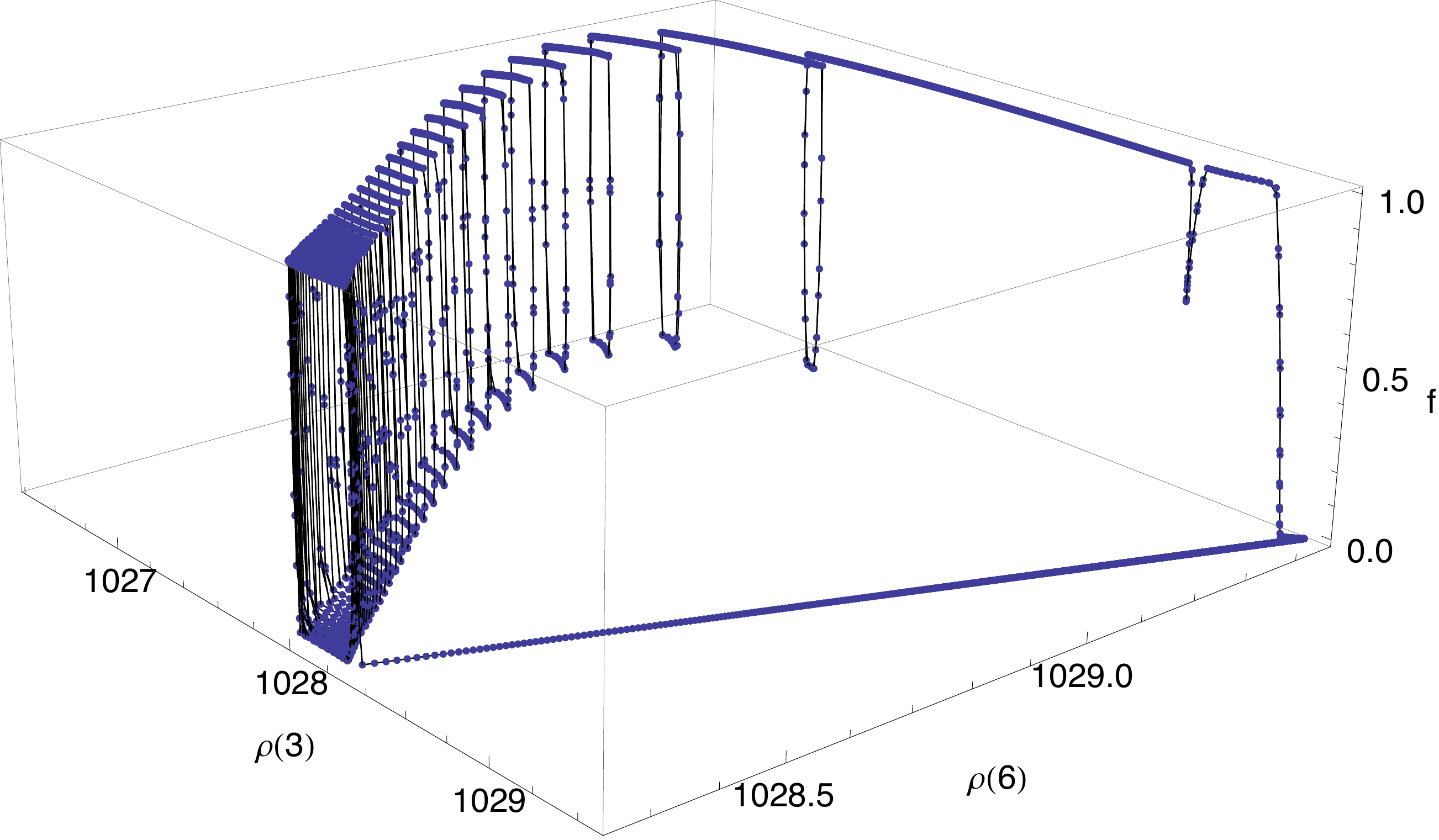}
\caption{A three dimensional phase space spanned by the densities of the polar boxes and sea ice extent. When the small amplitude high frequency oscillations touch the convective manifold, where $\rho_3=\rho_6$, an abrupt transition to a thermal mode occurs. 
}
\label{fig:3Dphase}
\end{center}
\end{figure}

High mixing rates in the convective phase remove the density anomalies that resulted in the convective instability. This allows the system to relax back to the small amplitude oscillations, which in the absence of sea ice would be a stable fixed point. The existence of sea ice destabilizes the HA equilibrium. For appropriate strengths of the salinity forcing, the small oscillations are interrupted when the system touches the convective manifold and gets forced away from the HA fixed point. A three-dimensional phase space that best captures the dynamical behavior of the system is spanned by the top and bottom polar densities and sea ice extent. Figure (\ref{fig:3Dphase}) shows one entire oscillation cycle, consisting of both the small and large amplitude components. 

\begin{figure}[htbp]
\begin{center}
	\includegraphics[width=3.5in]{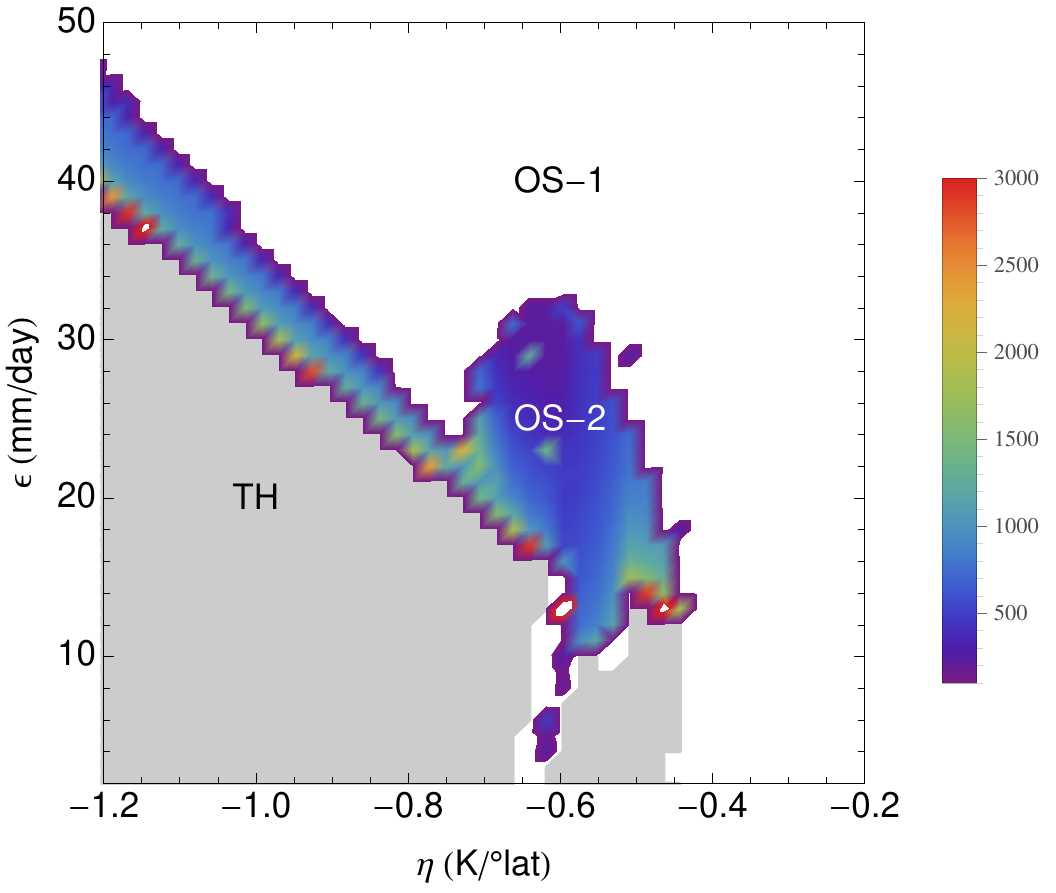}
\caption{
Model states in the two-dimensional forcing space. Larger magnitudes of $\eta$ imply steeper atmospheric temperature gradients.
}
\label{fig:forcingspace_SI}
\end{center}
\end{figure}

\begin{figure}[htbp]
\begin{center}
	\includegraphics[width=3.5in]{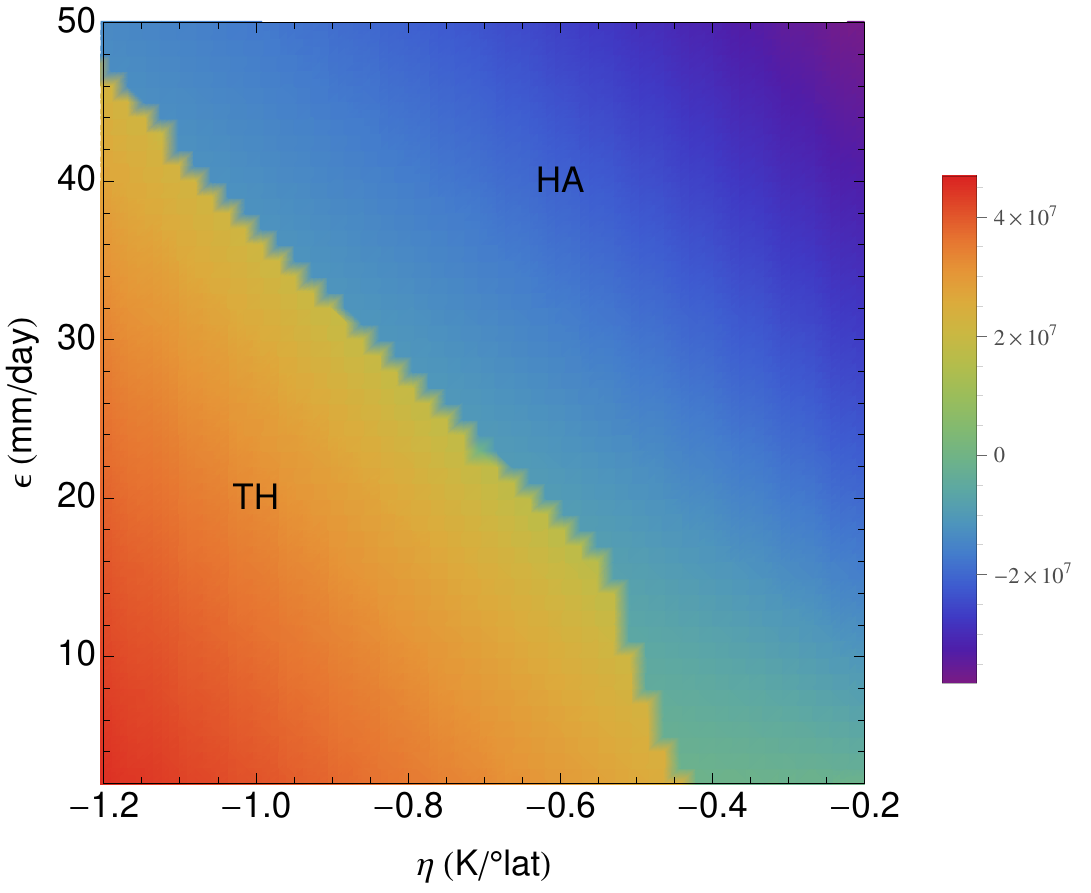}
\caption{
Only steady TH and HA states occur in the ocean model decoupled from sea ice.
}
\label{fig:forcingspace_noSI}
\end{center}
\end{figure}

The model's response is systematically examined in a large range of thermal and freshwater forcing values. Independent simulations with identical initial conditions are carried out in 50x50 pairs of forcing values. Detection of oscillations and/or steady states from each of the generated time series is automated. Equilibrium flux values, oscillation frequencies and amplitudes are recorded for each simulation. Figure (\ref{fig:forcingspace_SI}) shows the  states in the forcing space. For the set of simulations where sea ice is decoupled from the ocean model, the state space only has TH and HA states and no oscillations, figure (\ref{fig:forcingspace_noSI}). 

In both the sea ice coupled and decoupled cases, TH states occur when the relative strength of the thermal forcing is greater than the salinity forcing. HA states are only observed for extremely large salinity forcing where sea ice is coupled. OS-2 states in the sea ice coupled model occurs along the edge of TH and HA states where the relative strengths of the two forcings nearly balance each other. 

\subsection{Bifurcations}
\label{subsec:bifurcations}
\begin{figure}[htbp]
\begin{center}
	\includegraphics[width=3.5in]{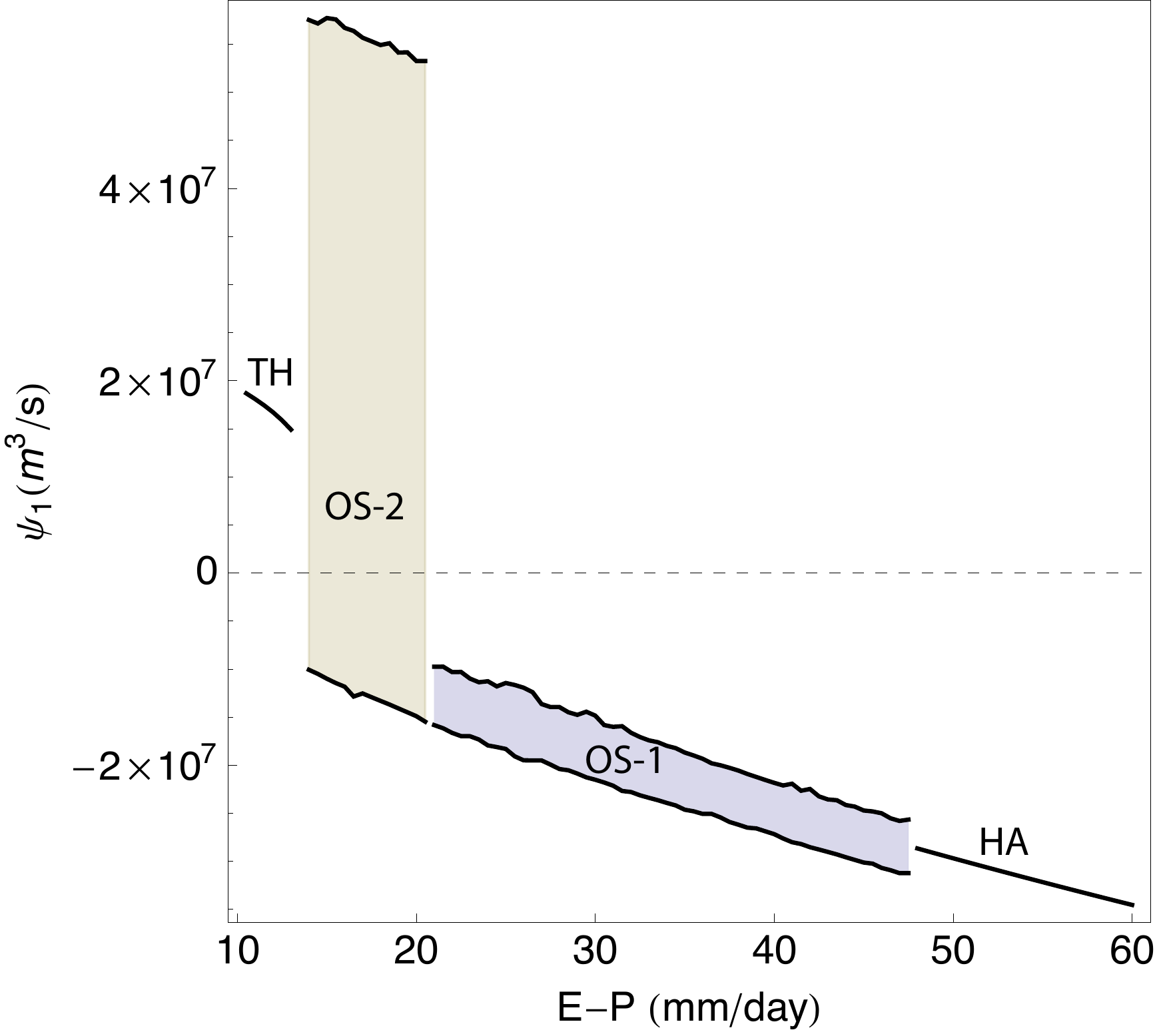}
\caption{Bifurcation diagram for slowly varying salinity forcing parameter $\epsilon$.}
\label{fig:bifurcations}
\end{center}
\end{figure}

The bifurcation structure is examined by slowly varying the salinity forcing parameter and observing the model's behavior, while keeping the thermal forcing parameter fixed. For every new parameter value the model is allowed to run for a sufficient length of time to let it come to equilibrium or sustained oscillations, as the case may be. Traversing through ascending values of $\epsilon$ shows all the four modes, TH, OS-2, OS-1, and HA in successive order, figure (\ref{fig:bifurcations}). 

\begin{figure}[htbp]
\begin{center}
	\includegraphics[width=3.5in]{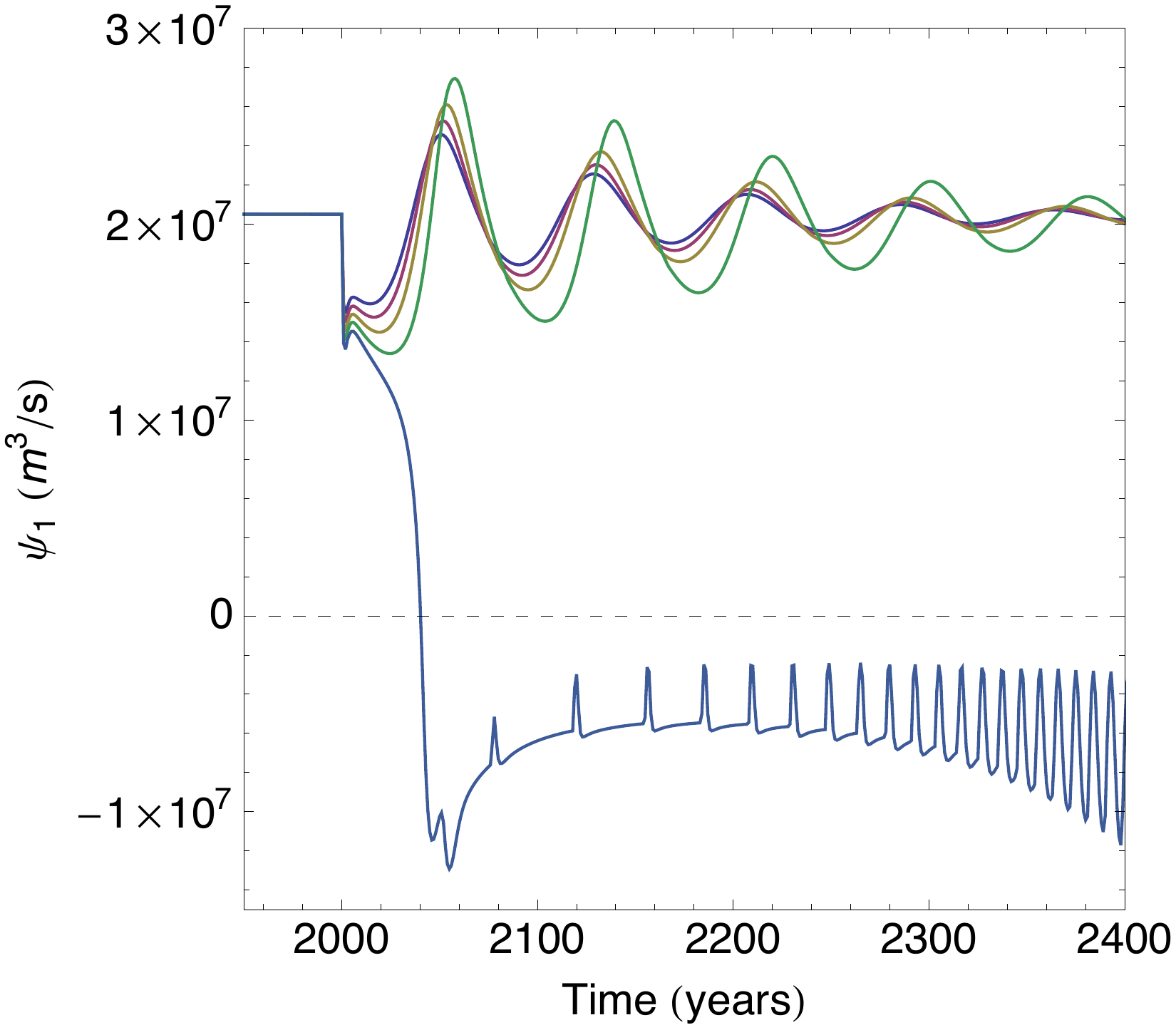}
\caption{Subcritical Hopf bifurcation from TH to OS-2 state. A series of progressively larger perturbations are applied to the steady TH state near the bifurcation point. Perturbations with amplitudes below a threshold value decay exponentially in time, while those larger are repelled to the haline regime of flow.}
\label{fig:bifurcations-TH-OS2}
\end{center}
\end{figure}

A steady TH state near the bifurcation point to OS-2 is given a series of perturbations by adding freshwater to the top polar box. Small magnitudes of perturbation decay exponentially and the system spirals and settles back into the steady state. If the perturbation is larger than a certain threshold, figure (\ref{fig:bifurcations-TH-OS2}), the trajectory diverges far, into the haline regime. This behavior is characteristic of subcritical Hopf bifurcations. As expected, the system exhibits hysteresis if the direction of varying $\epsilon$ is reversed. 

The OS-2 state is a mixed mode, with two oscillation components. The small amplitude high frequency component is near an unstable HA fixed point. It is the coupling of sea ice that makes this fixed point unstable, as without it the small oscillations do not occur. However, before the oscillations can settle on a limit cycle, they could encounter the convective manifold and get repelled towards a transient thermal mode. The abrupt jump and relaxation back to the haline mode is the large amplitude low frequency component of OS-2. Higher $\epsilon$ values push the limit cycles away from the convective manifold, so that the convective detour never takes place and the system persists in the small amplitude limit cycle mode, or OS-1. Thus the bifurcation from OS-2 to OS-1 is of the homoclinic type. 

The OS-1 to HA bifurcation occurs when sea ice completely fails to grow. As soon as sea can grow a small amplitude limit cycle appears, which suggests that a supercritical Hopf bifurcation has occurred. 

The dynamical properties of the ocean model by itself is similar to that of Stommel's box model in that it exhibits stable haline equilibrium modes and spirally stable thermal mode. \citet{ruddick_1996} prove that Stommel-like models cannot exhibit oscillations under steady forcing.

\section{Period dependence on parameters}

Oscillation timescales are determined by four main parameters - the ratio of the thermal to salinity forcing $\eta/\epsilon$, the growth-melt timescale for sea ice $\tau$,  the overturning flow rate constant $C$, and the maximum extent of sea ice. With the exception of the growth timescale, all other parameters varied significantly over the course of the last glacial period and the response of the sea ice-convection oscillator to those variations may have played a significant role in determining the climate of the north Atlantic. The model's response is analyzed for different values of these parameters.

\subsection{Ratio of thermal to salinity forcing magnitudes}
\begin{figure}[htbp]
\begin{center}
	\includegraphics[width=3in]{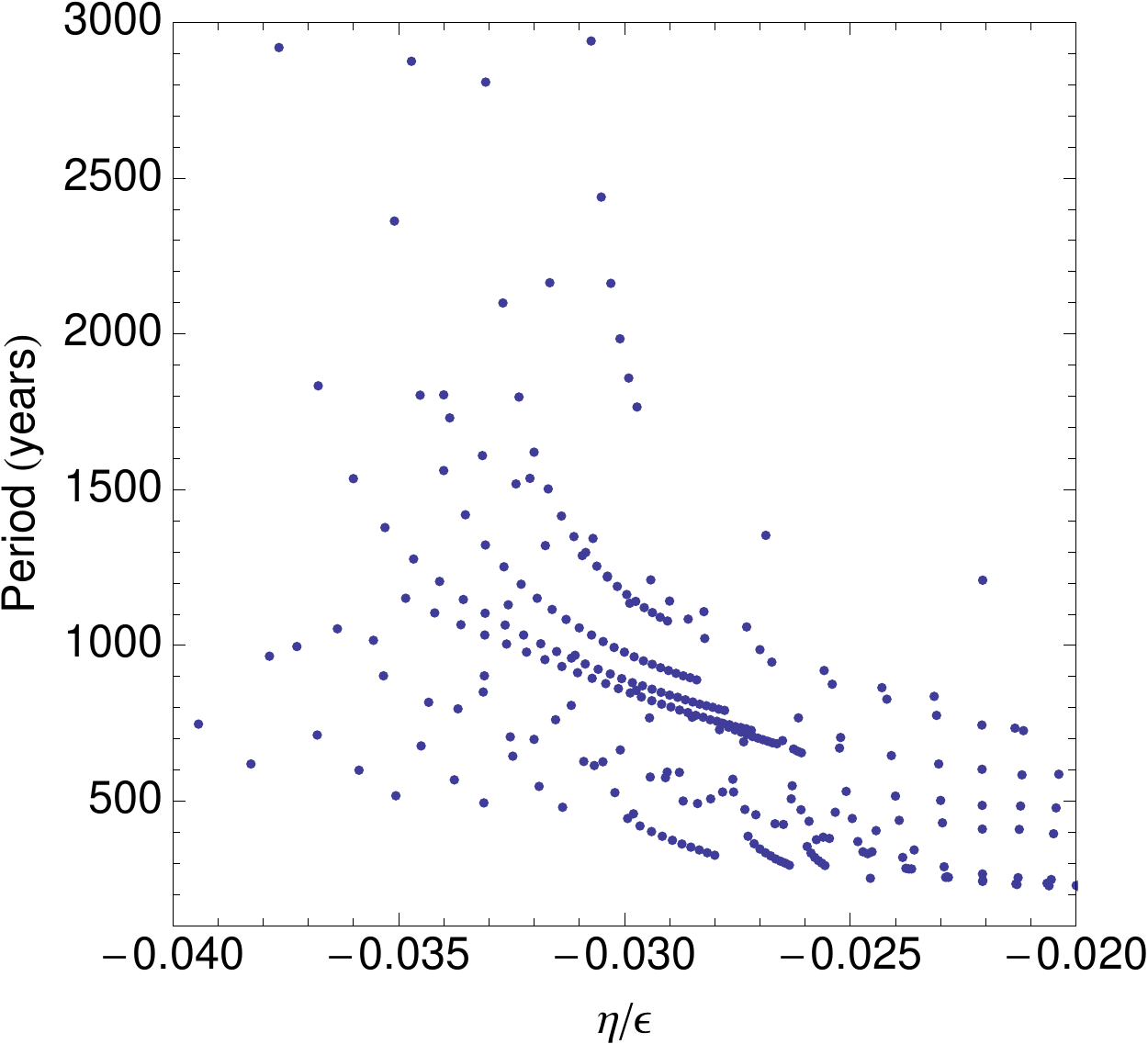}
\caption{Distribution of periods over the ratio of forcing strength parameters. Larger negative values of the ratio imply greater relative strength of thermal forcing.}
\label{fig:periods-etaemp}
\end{center}
\end{figure}

The mixed mode oscillations occur when the effects of thermal and salinity forcing nearly balance each other. Periods of oscillations have two components, the timescale for a convective instability to build up in the polar column, and the relaxation timescale for the system to return from the transient thermal mode. Strong atmospheric thermal gradients increase the attractive basin for thermal circulation modes and therefore tend to increase the relaxation timescale, which has the most pronounced effect on the oscillation period. Figure (\ref{fig:periods-etaemp}) tabulates the periods of all OS-2 states from figure (\ref{fig:forcingspace_SI}). It shows that the periods vary from a few hundred to a few thousand years.

While 1,500 years is the most commonly cited period for DO events, the proxy records show that their pacing varied quite significantly during the last 100,000 years. The most regular periodic occurrence of DO events is in the window of 30,000 and 50,000 years before present, which was also the time when high latitude summer insolation had the least amount of variation. In other time windows the pacing of DO events appear to be modulated by insolation variations. The distribution and duration of abrupt warming events during the last glacial period was probably shaped in large part by a combination of solar and glacial meltwater forcing.

\subsection{Timescale for sea ice growth/decay}
The growth and melt timescale for sea ice depends on a large number of physical conditions in the real world, such as the spatial and seasonal variations in near surface temperatures, ocean-atmosphere heat fluxes and strengths of ocean currents. In the model this timescale is prescribed to be in the 2-3 year range, which is largely an educated guess of how fast sea ice could grow over the Nordic seas under a static background climate similar to the last glacial period. During last glacial maximum the southern extent of winter sea ice was Iceland, whereas today it is near Svalbard.  The area of sea between Svalbard and Iceland is considerably larger than the present seasonal variations in sea ice extent. Considering the annual seasonal cycle of growth and melt the default timescale for sea ice is reasonable. 

\begin{figure}[htbp]
\begin{center}
	\includegraphics[width=3in]{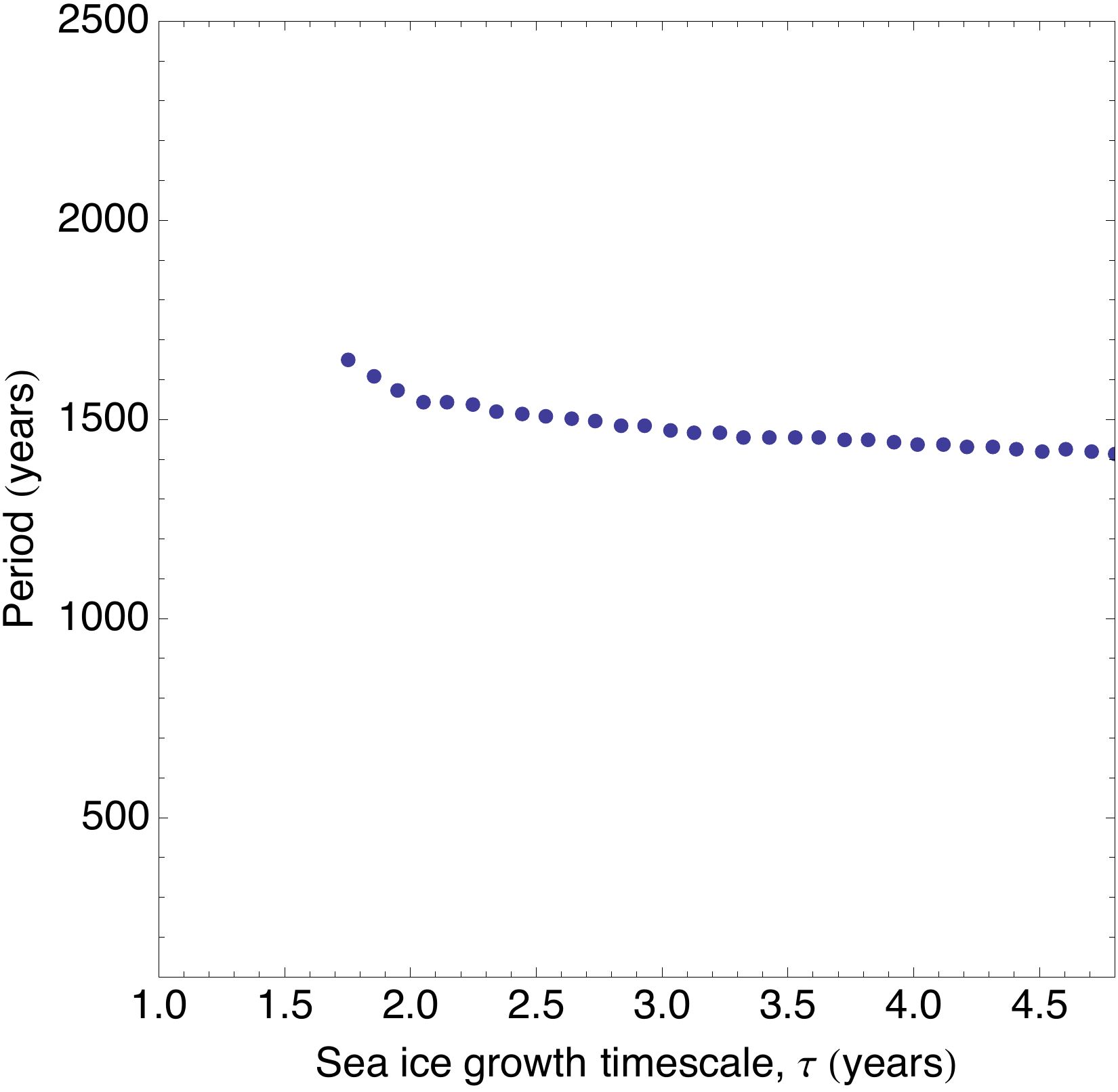}
\caption{Distribution of periods over the varying timescales of sea ice growth. No oscillations occur if $\tau<$ 2 years. Above this value the sensitivity of periods to the growth timescale is low. }
\label{fig:periods-tauSI}
\end{center}
\end{figure}

With all other parameters fixed at their default values the model is run for a range of sea ice timescales. An abrupt transition occurs around a value of 1.5 years, above which oscillations appear and approach periods of about 1,400 years (figure \ref{fig:periods-tauSI}). The threshold occurs where the insulating effect of sea ice in trapping heat in the ocean is matched by the dissipative heat loss by conduction and advection. The sensitivity for values of $\tau$ above this threshold is low as only the high frequency component is affected and not the longer timescale relaxation mechanism.

\subsection{Advective flux rate constant}
\begin{figure}[htbp]
\begin{center}
	\includegraphics[width=3in]{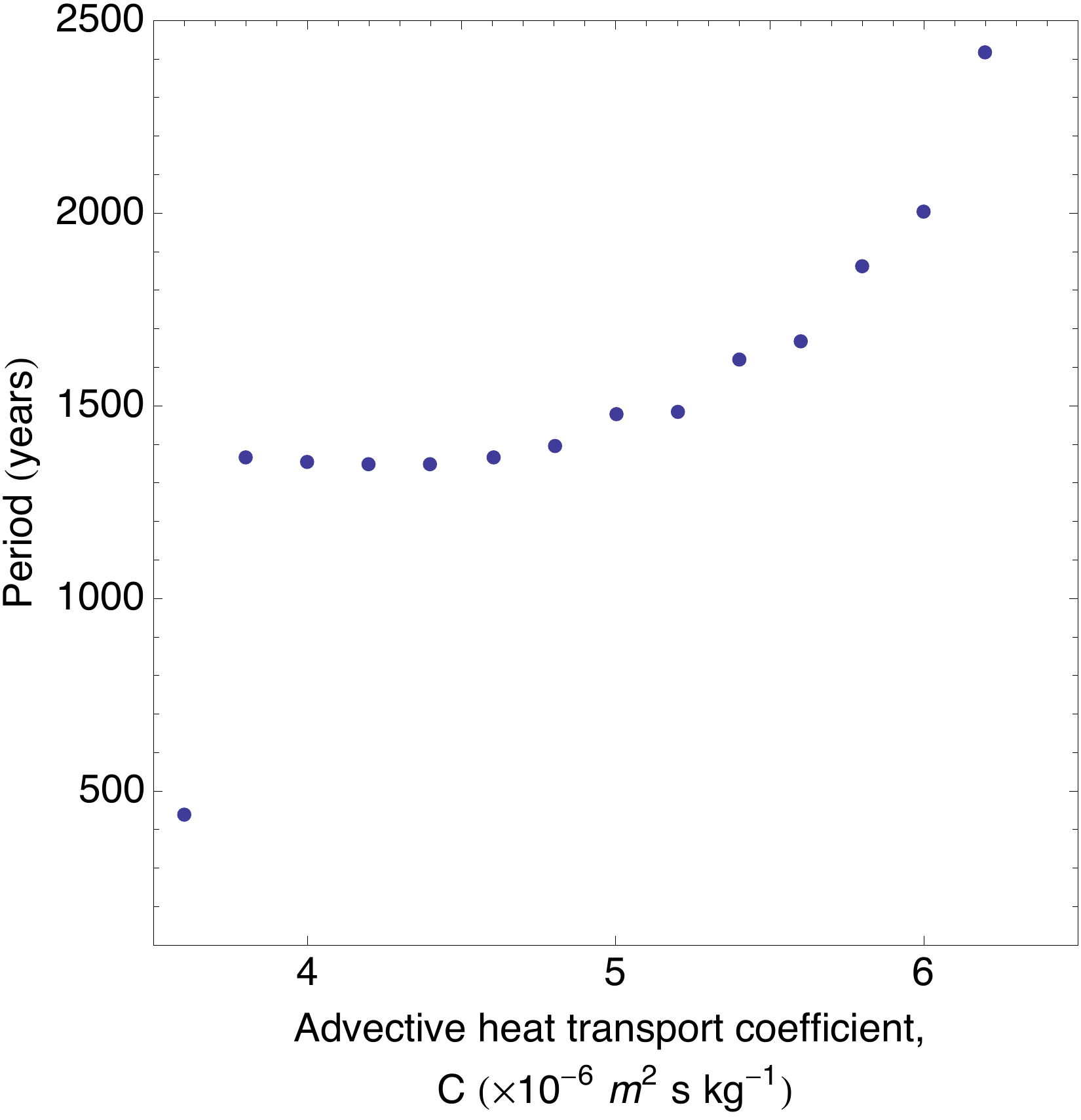}
\caption{Distribution of periods over the strength of advective flow rate.}
\label{fig:periods-C}
\end{center}
\end{figure}

The parameter $C$ prescribes the meridional flow rate per unit pressure gradient and its default value is based on current flow rates in the North Atlantic \citep{verdiere2006}. However, this rate was likely very different between glacial and interglacial times \citep{okazaki2010}. In the context of a relaxation oscillator, especially one where the flow is conserved, the rate of flow would influence timescales of both instability build up and relaxation. For small flow rates the heat build up in the ocean depths would fail to bring about convective instability, whereas high flow rates would equalize density anomalies efficiently and prevent oscillations. As a result the model should exhibit oscillations only within a range of flow rates, as observed in simulations over a range of flow rates (figure \ref{fig:periods-C}). 

\subsection{Maximum extent of sea ice}
The southern extent of sea ice in the north Atlantic is sensitive the background climatic state, glacial or interglacial, as evidenced by proxy data measuring sea ice extent \citep{sarnthein_2003}. However among the other parameters that control the periodicity of the system, the sea ice extent is most constrained due to the positioning of land masses and structure of ocean currents in the north Atlantic. Could it be possible that the characteristic 1,500 year period of DO events is due to the size of the ocean basin effected by sea ice? 

\begin{figure}[htbp]
\begin{center}
	\includegraphics[width=3in]{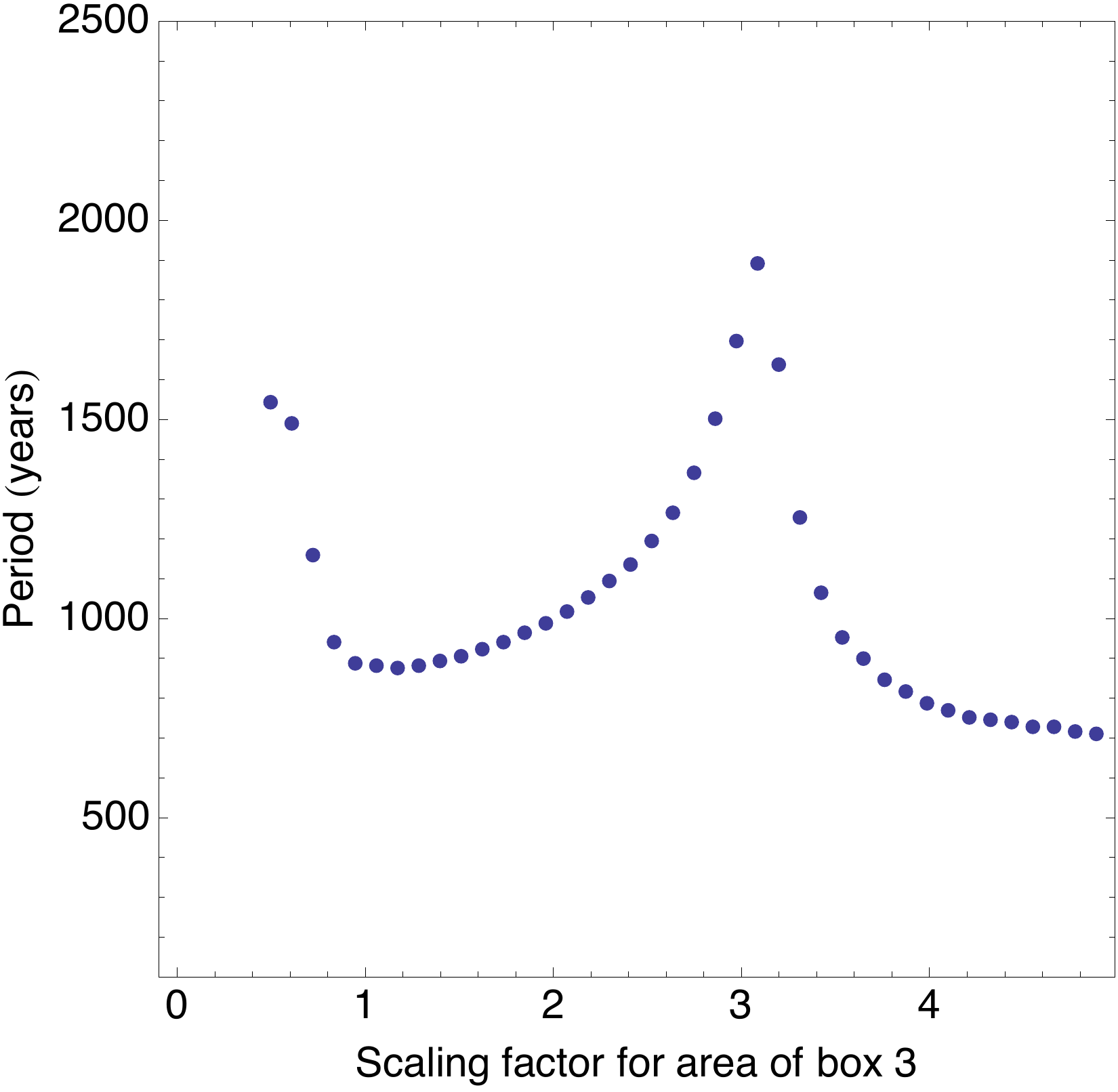}
\caption{
Oscillation periods for different polar areas. The horizontal axis is the scaling factor of the default polar area. 
}
\label{fig:periods-A3}
\end{center}
\end{figure}

To test this hypothesis the area of the top polar box is varied between factors of 0.5 and 5 of its default area, and oscillation periods observed in this range. Periods show high sensitivity to the polar area, which is also a measure of the maximum extent of sea ice. A peak in the periods is observed around a factor of 3, figure (\ref{fig:periods-A3}). The north Atlantic oscillator system in the real world could also have a similar non-linear sensitivity to the extent of sea ice. This would make certain oscillation periods favorable within certain windows of sea ice variance, and very different periods in other windows, as may have been the case between glacial and interglacial climates.

\section{Modulation by freshwater forcing}

The addition of freshwater to the ocean surface due to catastrophic collapse of ice sheets has been speculated to be the trigger for DO events. Simulations with intermediate complexity models do show that freshwater additions in the North Atlantic can abruptly destabilize and enhance the meridional circulation \citep{rahmstorf_2002,hofmann_rahmstorf_2009}. The higher complexity models, however, do not seem to capture an internal climatic oscillator and consequently the model responses do not exhibit periodic behavior in the absence of periodic freshwater forcing. Typically in these model studies every DO event needs to be triggered by a separate freshwater pulse, or an underlying freshwater periodicity imposed on the system. The climate record, however, suggests an intrinsic periodicity within the climate system that was also sensitive to freshwater pulses. 

\begin{figure}[htbp]
\begin{center}
	\includegraphics[width=3.5in]{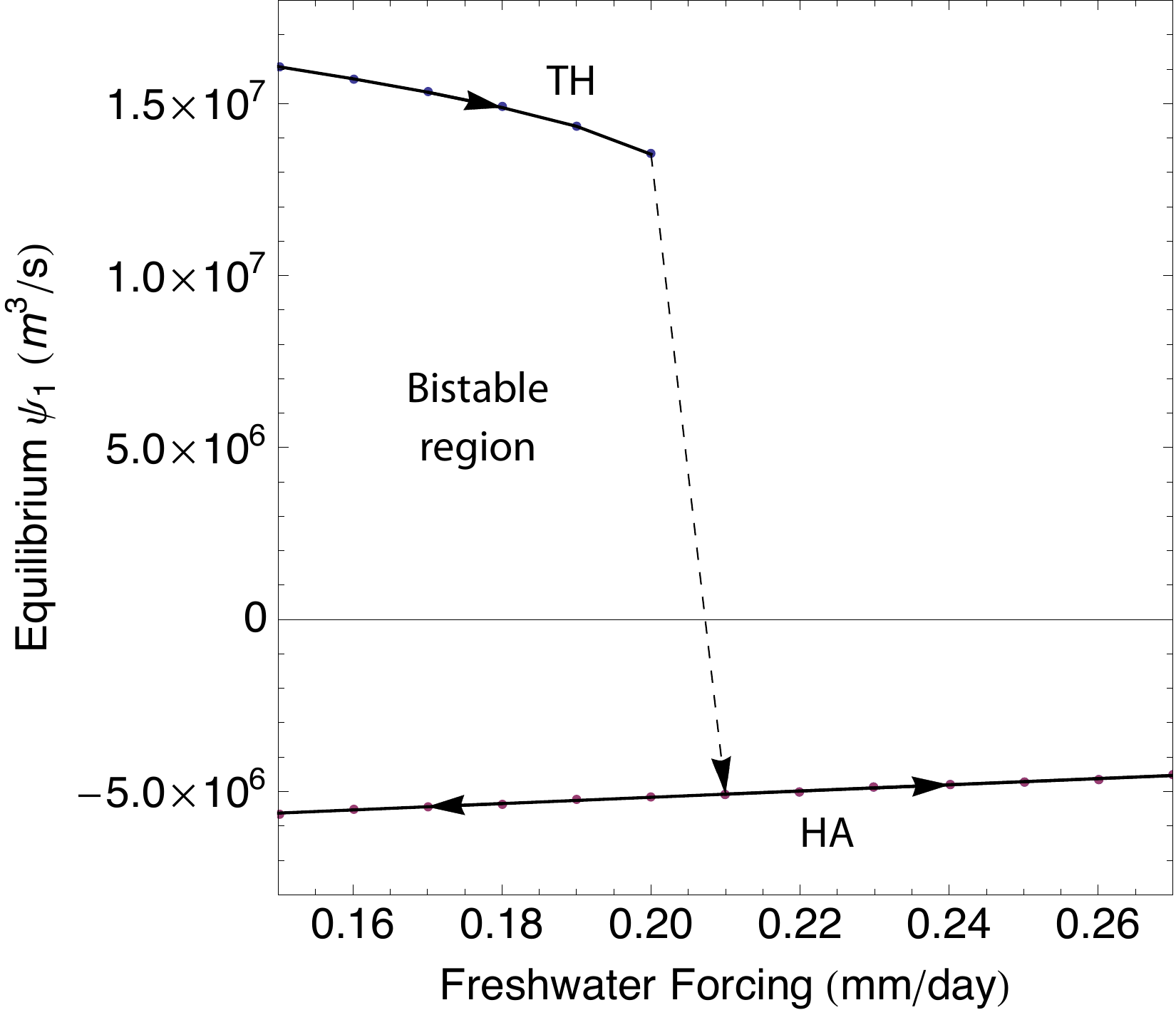}
\caption{
Steady states of the ocean box model decoupled from sea ice for varying rates of freshwater additions to the surface. A hysteresis is observed when the direction of freshwater additions is reversed. 
}
\label{fig:fwf-hysteresis}
\end{center}
\end{figure}

The effect of freshwater on the circulation state alone is examined by adding incremental amounts of freshwater to box 1, with no sea ice coupling on the polar box. The freshwater forcing is varied slowly, allowing the model to come to equilibrium in each step. A hysteresis is observed when the variation in freshwater additions is reversed, showing a bi-stable region and a stable HA mode, figure (\ref{fig:fwf-hysteresis}). In the absence of sea ice there is no mechanism for the circulation to undergo relaxation oscillations and consequently those are not observed.

The effect of Heinrich events is simulated by constructing a corresponding freshwater signal and forcing the sea ice coupled model with it. The pulses are spaced in intervals of 7,500 years and are represented by a Gaussian function. A smaller magnitude skewed Gaussian is added in between pulses to represent increasing basal melting from ice sheets.

\begin{figure}[htbp]
\begin{center}
	\includegraphics[width=5in]{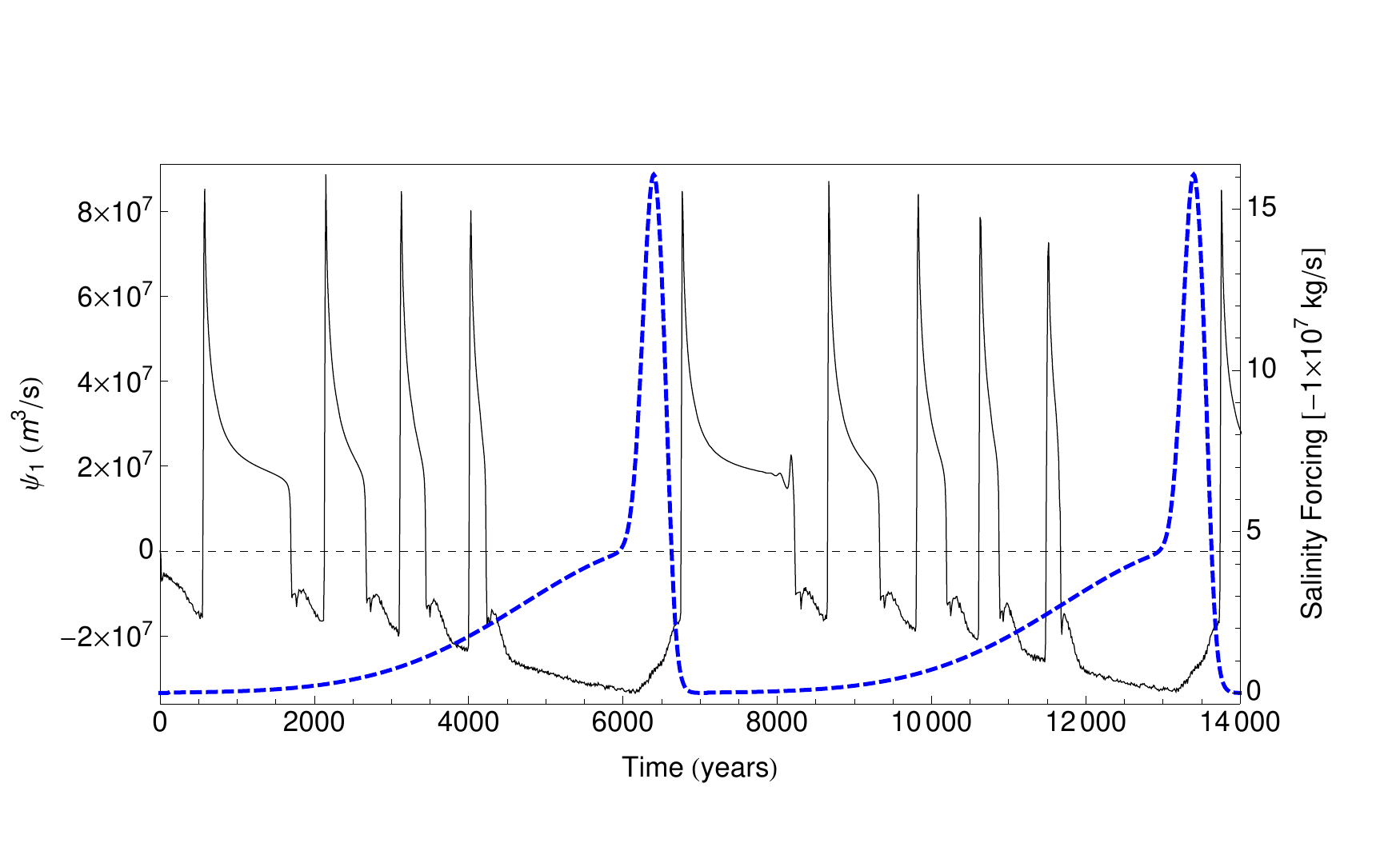}
\caption{
Forcing by Heinrich events (blue curve) can modulate the response of the sea ice-convective oscillator into patterns that are similar to the temperature fluctuations recorded in ice core proxies.
}
\label{fig:HR_sim}
\end{center}
\end{figure}

The otherwise periodic response of the model is modulated by the freshwater forcing for Heinrich events, figure (\ref{fig:HR_sim}). Each Heinrich event triggers a series of progressively weaker abrupt warmings, similar to what is observed in the climate record. The gradual addition of freshwater to the surface in between Heinrich events decreases the meridional pressure gradient and results in weaker transient thermal modes of circulation. The freshwater pulse causes the poleward circulation to abruptly cease, which initiates sea ice growth, which in turn causes a resumption of convection and poleward circulation.

Variation of high latitude summer insolation during the last glacial period was least in the window of 35,000 to 50,000 years before present. This is the same time window where DO events occurred with the regularity simulated with the model, possibly due to minimal influence of insolation forcing in that period. Speculating further it could be stated that DO events were most pronounced during the glacial period, in comparison to the Holocene, because the background circulation was in a weaker state. A combination of solar and surface freshwater forcing could explain the climatic fluctuations during the last 100,000 years, which is out of the scope of this paper and largely a diagnosis of the system.

\section{Discussion and conclusions}

A simple dynamical model is used to show that sea ice and convection in the north Atlantic could interact to produce millennial scale quasi-periodic fluctuations. In the idealized model described here,  the fluctuations are stable periodic orbits under a large range of climatic forcing. The timescale of model oscillations and the temporal pattern of variability is very similar to that of DO events and Bond Cycles. Time varying freshwater anomalies due to ice sheet discharge and variations in obliquity can modulate the intrinsic periodic response of the sea ice-convection oscillator. This oscillator could be at the heart of the millennial scale climatic variability known to have occurred during the last 100,000 years.

The oscillation is due a periodic build up of convective instability in the polar ocean. Sea ice cover, when present, insulates the ocean from losing heat to the atmosphere. However, over repeated cycles of sea ice advance and retreat the upper mixed oceanic layer undergoes a net loss of heat due to the large heat ventilation during retreat phases. The gradual increase in mixed layer density leads to convective instability which initiates an abrupt increase in vertical mixing in the polar water column. The high mixing rate removes the density anomalies and allows the circulation to relax back to its preferred state of small amplitude oscillations in sea ice advance and retreat. From here on the net positive heat loss from the mixed layer helps build up the convective anomaly and repeat the process in a self-sustained manner. 

Dynamically, the oscillations are composed of two components operating on a decadal and millennial timescale respectively. The presence of sea ice is responsible for the small amplitude oscillations, which are not observed in the ocean model decoupled from sea ice. These oscillations approach a limit cycle, but may be interrupted by the trajectories running into the convective manifold which force the system away from the unstable HA fixed point towards a transient thermal mode. The onset of convection brings about a high mixing rate which gradually removes the density anomalies and allow the circulation to relax back to the haline mode small oscillations. The mixed mode character of the sea ice oscillator means that there are at least three principal dynamical variables that are necessary ingredients. These are likely the horizontal and vertical density gradients in the north Atlantic ocean, and the extent of sea ice. Could it be that the well observed Atlantic Multi-decadal Oscillations are the small amplitude component of a mixed mode oscillation of the sea ice-convection system?

The period of oscillations in the model are dependent on the climatic forcing as well as physical attributes of the system. Sea ice extent is geophysically and climatologically limited within the corridor of Greenland-Icelandic-Norwegian seas and this could set the intrinsic timescale of 1,500 years. Freshwater pulses mimicking Heinrich events modulate the periodic response into repeating groups of successively weaker and shorter fluctuations. This response is similar to the Greenland temperature proxy signal in the window of 35,000 to 50,000 years before present. In this time window the summer insolation at high latitudes had very low variability as compared to the rest of the glacial period and thus surface freshwater forcing may have been the dominating influence. 

A box model, in spite of its simplicity, is an effective tool in identifying the fundamental mechanisms and ingredients behind the climatic phenomenon addressed in this paper. The persistent fluctuations occurring through both glacial and interglacial periods hint at a robust oscillation mechanism for which the sea ice-convection oscillator is a suitable candidate. New mathematical techniques to study non-smooth dynamical systems such as this one are needed for further exploration of the dynamics of this climate oscillator. 

\section*{Acknowledgements}
I would like to thank Chris Jones and John Bane for advising this research project, Mary Lou Zeeman and Andrew Roberts for helpful inputs and suggestions at various points. I would also like to thank NSF and the Mathematics and Climate Research Network for support.

\section*{References}

\bibliography{references}

\end{document}